\documentclass[epj,nopacs]{svjour}
\usepackage{graphics}
\usepackage{epsfig}
\begin{document}
\def\NJ{N_{\J}}
\def\J{J/\psi}
\def\Npart{N_{part}}
\def\Nc{N_c}
\def\Nch{N_{ch}}
\def\Ncbar{N_{\bar c}}
\def\ccbar{c \bar c}
\def\Nccbar{N_{\ccbar}}

\title{Quarkonium formation in statistical and kinetic models}
\author{R. L. Thews 
}                     
%
%
\institute{Department of Physics, University of Arizona,
Tucson AZ 85721 USA}
\date{Received: date / Revised version: date}
%
\abstract{
I review the present status of two related models addressing
scenarios in which formation of heavy quarkonium states in 
high energy heavy ion collisions proceed via ``off-diagonal" combinations
of a quark and an antiquark.  The physical process involved belongs to
a general class of quark ``recombination", although technically the
recombining quarks here were never previously bound in a quarkonium state.
Features of these processes relevant as a signature of color
deconfinement are discussed. 
\PACS{
      {PACS-key}{discribing text of that key}   \and
      {PACS-key}{discribing text of that key}
     } 
} 
\maketitle
\section{Introduction}
\label{intro}
The original idea of Matsui and Satz \cite{matsuisatz} predicted a 
suppression of $\J$ produced in heavy ion collisions as a result of the
expected screening of the color force above the deconfinement phase transition.
The prediction of suppression follows from the expectation 
 that the eventual hadronization of the deconfined
charm quarks is preferentially with light up and 
down quarks, since generally only one $\ccbar$ pair is produced in a given
collision.  

Several years ago, it was pointed out that the suppression scenario could be
altered in nuclear collisions at collider energies \cite{stathad,kinetic}.
At sufficiently high
energy, multiple pairs of heavy quarks will be produced in a single 
nucleus-nucleus collision. Then it may be possible for a given heavy quark
to form a heavy quarkonium hadron by combining with a heavy antiquark which
originated from a different initial production process.  I will refer
to such combinations as ``off-diagonal" pairs.  The probability to form
heavy quarkonium will of course depend on the physics of the interaction and
also the nature of the medium in which it occurs.  However, one can predict 
a few simple properties of this formation process based on general considerations.

We consider scenarios in which the formation of $\J$ is allowed to
proceed through any combination consisting of one of the $\Nc$ charm quarks with
one of the $\Ncbar$ anticharm quarks which result from the initial production
of $\Nccbar$ pairs in a central heavy ion collision.  For a given charm quark,
one expects then that the probability $\cal{P}$ to form a $\J$ is just proportional
to the number of available anticharm quarks relative to the number of
light antiquarks.
\begin{equation}
{\cal{P}}_{c \rightarrow \J}
 \propto \Ncbar / N_{\bar u + \bar d + \bar s} \approx \Nccbar / \Nch,
\end{equation}
where we normalize the number of light antiquarks by the number of
produced charged hadrons.
Since this probability is generally very small, one can simply multiply by the
number of available charm quarks $\Nc$ to obtain the total number of
$\J$ expected in a given event.

\begin{equation}
\NJ \propto {\Nccbar}^2 / N_{ch},
\end{equation}
where the use of the initial values $\Nccbar = \Nc = \Ncbar$ is again justified
by the relatively small number of $\J$ formed.
For an ensemble of events, we calculate 
the average number of $\J$ per event
from the average value of ${\Nccbar}^2$, and neglect fluctuations
in $\Nch$.
\begin{equation}
\label{quadratic}
<\NJ> = \lambda (<\Nccbar>+1)<\Nccbar> / \Nch,
\end{equation}
where we place all dynamical dependence in the parameter $\lambda$.
The resulting quadratic dependence on $<\Nccbar>$ provides a unique signature
which must at some high energy become dominant over production via a superposition
of independent diagonal $\ccbar$ pairs.

Initial estimates of $\Nccbar \approx 10$ for central Au-Au collisions 
at RHIC used extrapolations of cross section measurements at lower energy \cite{vogt}.
More recently there are measurements at RHIC based on high-transverse momentum
electrons by PHENIX and also reconstructed D-mesons by STAR which imply larger
numbers.  Central values of these measurements lead to  $\Nccbar \approx$ 20 
(PHENIX)\cite{phenixcharm} or 
40 (STAR)\cite{starcharm}, 
with relatively large experimental uncertainties which leave the two
measurements consistent.  In the following estimates for $\J$ we explore
this entire range of initial $\Nccbar$.

\section{Statistical Hadronization}
\label{stathad}
This model
was motivated by the success of predictions for the
relative abundances of light hadrons produced in high energy 
heavy ion interactions in
terms of a hadron gas in chemical and thermal equilibrium.
Such fits, however, underpredict the abundances 
of hadrons containing charm quarks.  This can be understood in terms of the
long time scales required to approach chemical equilibrium for heavy quarks,
starting from the large number of charm quarks produced via hard processes
during the initial stages of the collision.
The original formulation \cite{stathad}
of the statistical hadronization model
for hadrons containing charm quarks assumes that at hadronization the charm 
quarks are distributed into
hadrons according to chemical equilibrium, but adjusted by a factor
$\gamma_c$ which accounts for oversaturation of charm.  One power of
this factor multiplies
a given thermal hadron population for each charm or anticharm
quark contained in the hadron.  Thus the relative abundance of
$\J$ to that of D mesons, for example, will be enhanced in this model.
The enhancement factor is determined by conservation of charm, again
using the time scale argument to justify neglecting pair production
or annihilation before hadronization.

\begin{equation}
\label{charmcons}
\Nccbar = {1\over 2}\gamma_c N_{open charm} + {\gamma_c}^2 N_{hidden charm},
\end{equation}
where $N_{open charm}$ and $N_{hidden charm}$ are calculated in the thermal
equilibrium model grand canonical ensemble. (For peripheral collisions the
total particle numbers are not sufficiently large and one must calculate in the
canonical ensemble.)  It was first shown numerically in Ref. \cite{thewsbielefeld},
and later motivated in Ref. \cite{kostyuk} that the canonical correction effect
is equivalent to directly using the grand canonical value for $\gamma_c$ in
the ensemble average according to Eq. \ref{quadratic}.
The hidden charm term is negligible for all cases of present interest, and one
finds that $\gamma_c$ is directly proportional to $\Nccbar$.  Then one can express
the number of $\J$ at hadronization as

\begin{equation}
\NJ = {\gamma_c}^2 n_{\J}V,
\end{equation}
where n and V are the number density and volume appropriate to the relevant
hadronization region.  Insertion of the expression for $\gamma_c$ from Eq. \ref{charmcons}
leads to an expression which has the form expected in Eq. \ref{quadratic}, with
\begin{equation}
\lambda = {4 n_{ch} n_{\J} \over (n_{open})^2}.
\end{equation}
(Note that the factor of $\Nch$ appears  due to  replacing the one remaining power of V 
by the ratio of total number to density for charged particles.)
\subsection{Comparison with RHIC measurements}
There has been one measurement at RHIC by PHENIX for $\J$ production in Au-Au 
interactions at 200 GeV \cite{phenixjpsi}.  The data was analyzed in three
centrality regions, but due to limited statistics the uncertainties were 
quite large.  Also, the most central data leads to only an upper limit.  Two separate
groups \cite{andronic,kostyuk} have applied the statistical hadronization model in this case.  Both have
found general agreement with the data, which involves rapidity densities at y = 0 
rather than total yields.  (Only the two more central data points can be used, since
measurements of the relative yields of $\J$ and $\psi^{\prime}$ are consistent with
the thermal model only in this region.)  However, the charm production cross sections
used in these calculations were different (390 $\mu$b vs. 650 $\mu$b), which 
would imply a difference in predictions of almost a factor of 3, all other effects being equal.
Although the thermal parameters appear to be compatible, the extraction of a
rapidity-density volume parameter evidently is different in these two approaches.

\subsection{Centrality dependence}
It is now conventional in heavy ion collisions to parameterize the centrality of
the collision in terms of the number of nucleon participants, $\Npart$.  The
point-like process in which $\ccbar$ pairs are produced then leads to a 4/3 power
law behavior of $\Nccbar$.

\begin{equation}
\Nccbar = \Nccbar (0) \left({\Npart \over2A}\right)^{4/3},
\end{equation}
which is normalized by the maximum number $\Nccbar(0)$ at impact parameter b=0 where 
$\Npart \approx$ 2A.

One also requires the centrality dependence of $\Nch$, for which we
parameterize $\Nch = a (\Npart)^{1+\Delta}$, where a is a normalization factor 
and $\Delta$, which depends on the production process for charged particles, 
will be varied.  Then the centrality dependence of $\NJ$ is determined
by substitution into Eq. \ref{quadratic}.  One generally normalizes the experimental yield
of $\J$ by either the number of binary collisions (equivalently $\Nccbar$),
or the number of nucleon participants $\Npart$.  For the first normalization
choice, one obtains

\begin{equation}
{\NJ \over\Nccbar} = {\lambda \over a} \left( (\Npart)^{{1 \over 3}
-\Delta}  +  (\Npart)^{-1-\Delta}\right).
\end{equation}

This combination of two power-law terms in $\Npart$ which differ by
${4 \over 3}$ is obviously due to the combination of quadratic and linear
terms in $\Nccbar$ for $\NJ$.  ${\NJ \over \Nccbar}$ will have a 
minimum (for $\Delta < {1 \over3}$) at $\Npart = N_{min} \equiv  
2A \left[ {(1+\Delta) \over {({1 \over 3} - \Delta) \Nccbar (0)}}\right]^{3/4}$.  
The sharpness of 
the minimum can be characterized by the ratio

R = ${\NJ \over \Nccbar}(N_{min}) / {\NJ \over\Nccbar}(2A)$, 
with the result
\begin{equation}
R = {4({1 \over 3} - \Delta)^{ (\Delta-1/4)} \over{3 (\Nccbar(0)+1)}} 
\left[ {\Nccbar(0) \over {1+\Delta}}\right]^{{3 \over 4}(1+\Delta)}
\end{equation}

\begin{figure}
\vskip 1.0cm
\resizebox{0.5\textwidth}{!}{%
  \includegraphics{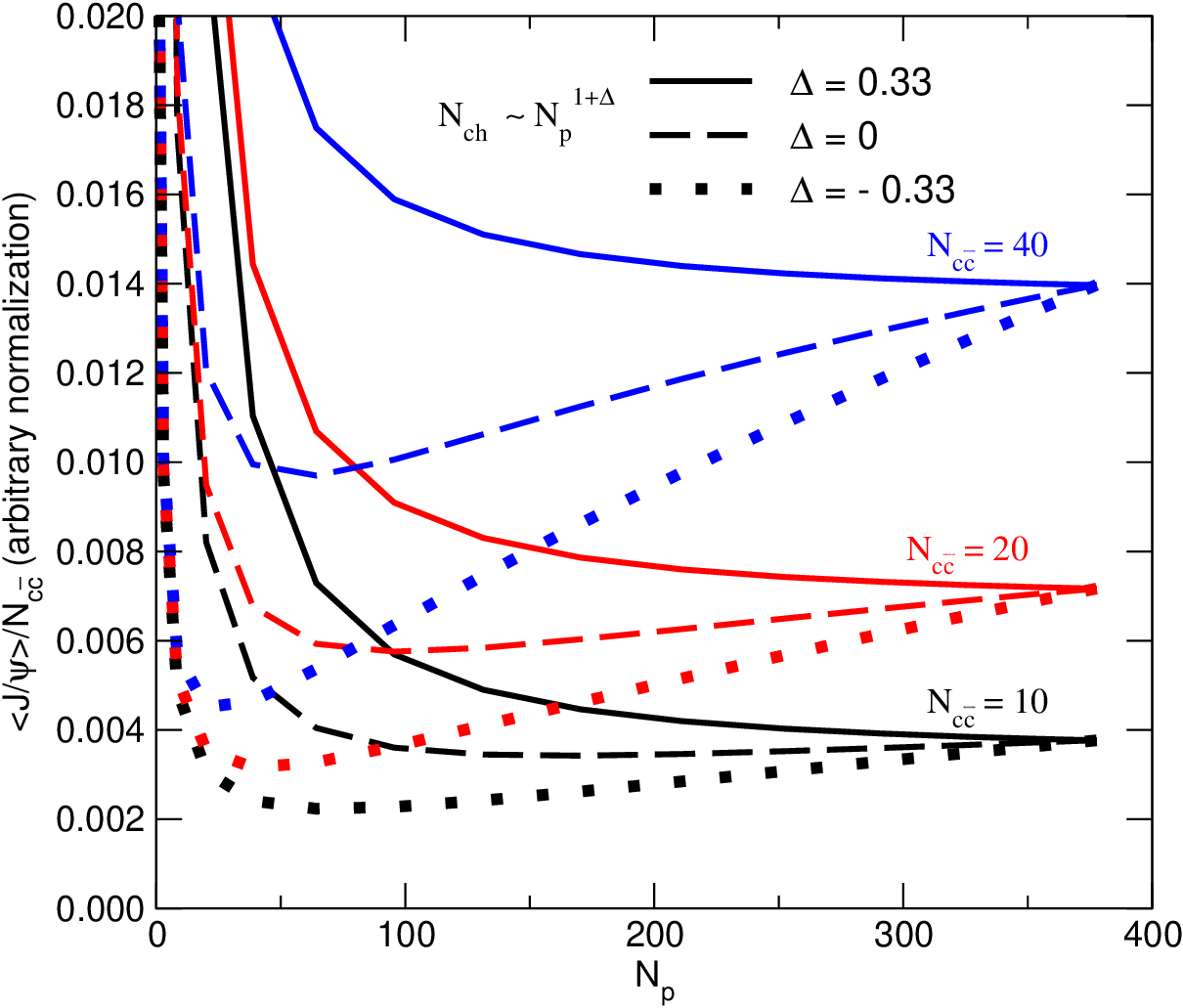}
}
\caption[]{Centrality dependence of binary-scaled $\J$ formation
via statistical hadronization.
}
\label{stathadbinary}
\end{figure}
These features are shown in Fig. \ref{stathadbinary} for a range of
$\Delta$ and $\Nccbar(0)$.  Aside from the curves for $\Delta = {1 \over 3}$ which 
are constant for large $\Npart$, all of the minimum points are at
relatively low values of $\Npart$, in the region where the statistical
hadronization cannot be applied.
\begin{figure}
\vskip 1.0 cm
\resizebox{0.5\textwidth}{!}{%
  \includegraphics{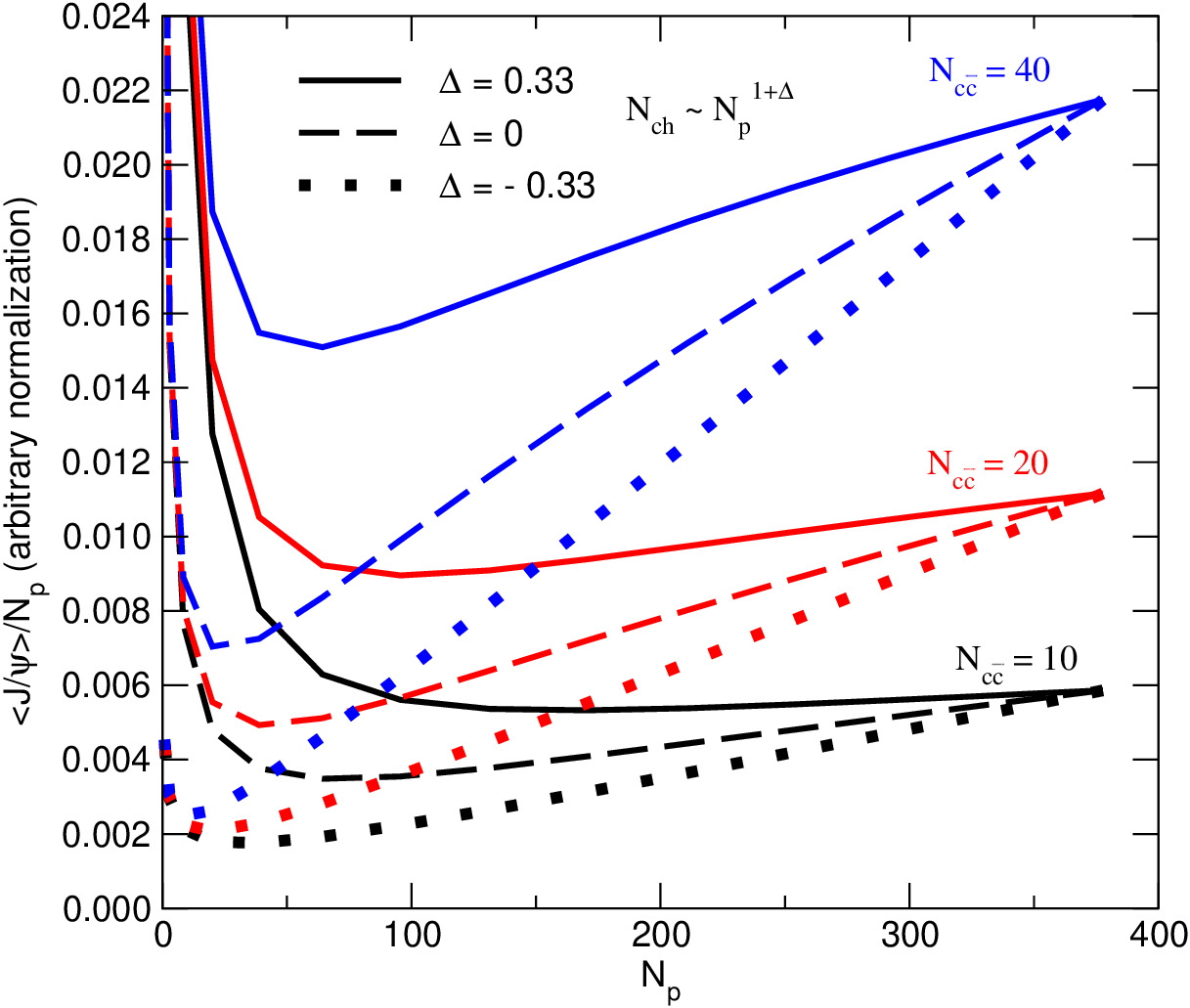}
}
\caption[]{Centrality dependence of participant-scaled $\J$ formation
via statistical hadronization.
}
\label{stathadparticipant}
\end{figure}

Figure \ref{stathadparticipant} shows the corresponding behavior for the
ratio ${\NJ \over\Npart}$, which can be obtained by making the
substitution $\Delta \rightarrow \Delta - {1 \over 3}$.  The same
general behavior is seen, but the sharpness of the approach to
the minimum value is enhanced, especially for the largest values
of $\Nccbar (0)$.  The real test of the predicted centrality
behavior requires data on both  the magnitude of the
initial charm production and the centrality dependence of $\Nch$. 
%
\section{Kinetic formation}
The kinetic model\cite{kinetic} describes a scenario in which the mobility of 
initially-produced charm quarks in a space-time region of
color deconfinement allow formation of quarkonium via ``off-diagonal" combinations
of quark and antiquark. The motivation for such a scenario in the case of
$\J$ formation has received support 
from recent lattice calculations of spectral functions.  These indicate that
$\J$ will exist in an environment at temperatures well above the deconfinement
transition \cite{lattice1,lattice2}.  The dominant formation process in this
scenario involves the capture of a quark and antiquark in a relative color octet
state into the color singlet $\J$ with the emission of a color octet gluon. This
reaction is just the inverse of the primary $\J$ dissociation process via collisions
with deconfined gluons \cite{kharzeevsatz}.  One can then follow the time evolution
of charm quark and charmonium numbers in a region of color deconfinement according
to a Boltzmann equation in which the formation and dissociation reactions compete.

\begin{equation}\label{eqkin}
\frac{d\NJ}{d\tau}=
  \lambda_{\mathrm{F}} N_c\, N_{\bar c }[V(\tau)]^{-1} -
    \lambda_{\mathrm{D}} \NJ\, \rho_g\,,
\end{equation}
with $\rho_g$ the number density of gluons.
The reactivity $\lambda$ is
the reaction rate $\langle \sigma v_{\mathrm{rel}} \rangle$
averaged over the momentum distribution of the initial
participants, i.e. $c$ and $\bar c$ for $\lambda_F$ and
$\J$ and $g$ for $\lambda_D$.
The gluon density is determined by the equilibrium value in the
QGP at each temperature, and the volume must be modeled according to the
expansion and cooling profiles of the interaction region.
 
This equation has an analytic solution in the case where the total number
of formed $\J$ is much smaller than the initial number of $\Nccbar$.

\begin{equation}
\NJ(\tau_f) = \epsilon(\tau_f) [\NJ(\tau_0) +
\Nccbar^2 \int_{\tau_0}^{\tau_f}
{\lambda_{\mathrm{F}}\, [V(\tau)\, \epsilon(\tau)]^{-1}\, d\tau}],
\label{eqbeta}
\end{equation}
where $\tau_f$ and $\tau_0$ are the final and initial times.
The function $\epsilon(\tau_f) = 
e^{-\int_{\tau_0}^{\tau_f}{\lambda_{\mathrm{D}}\, \rho_g\,
d\tau}}$
would be the suppression factor in this scenario if the
formation mechanism were neglected.

\begin{figure}
\vskip 1.0cm
\resizebox{0.45\textwidth}{!}{%
  \includegraphics{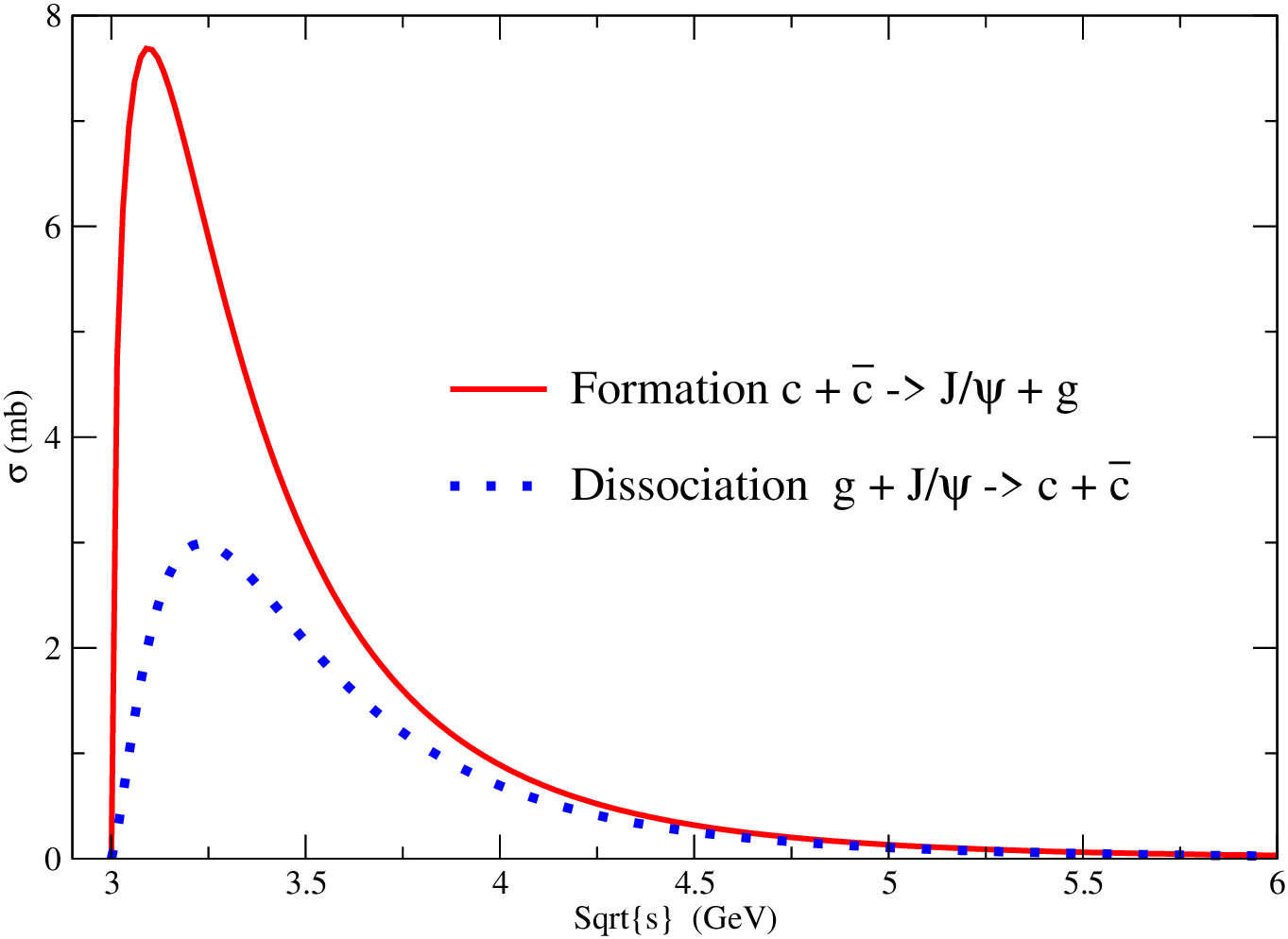}
}
\caption[]{Energy dependence of the OPE-inspired formation and dissociation
cross sections for $\J$.
}
\label{opecrosssections}
\end{figure}

One can readily see that $\NJ$ obeys the general properties present in 
Eq. \ref{quadratic}. However, the factor equivalent to system volume
V (in the statistical hadronization model) is time dependent and modified by a combination of factors involving the
interaction rates. Thus the centrality behavior  will depend on additional parameters.
The initial calculations \cite{thewsbielefeld} 
used the ratio of nucleon participants to participant
density to define a transverse area which defines the boundary of the
region of color deconfinement.  This is supplemented by longitudinal expansion
starting at an initial time $\tau_{0}$ = 0.5 fm (Transverse expansion was
initially neglected, but has been included in subsequent calculations \cite{thewssqm2003}.)
The expansion was taken to be isentropic, which determines the time evolution behavior  of
the temperature.  The initial value $T_{0}$ is taken as a parameter, and the
final $T_{f}$ is fixed at the hadronization point.  The reactivities $\lambda_{F}$ 
and $\lambda_{D}$ require specification of cross sections.  For $\sigma_{D}$ we 
use the ``OPE-inspired" model of gluon dissociation of deeply-bound heavy quarkonium
 \cite{peskin} \cite{kharzeevsatz}, which is related via detailed balance to the corresponding
$\sigma_{F}$.  These cross sections are shown in Fig. \ref{opecrosssections}.
One sees that they are peaked at low energy, and that $\sigma_{F} > \sigma_{D}$
due to the large binding energy.  

\begin{figure}
\vskip 1.0cm
\resizebox{0.45\textwidth}{!}{%
  \includegraphics{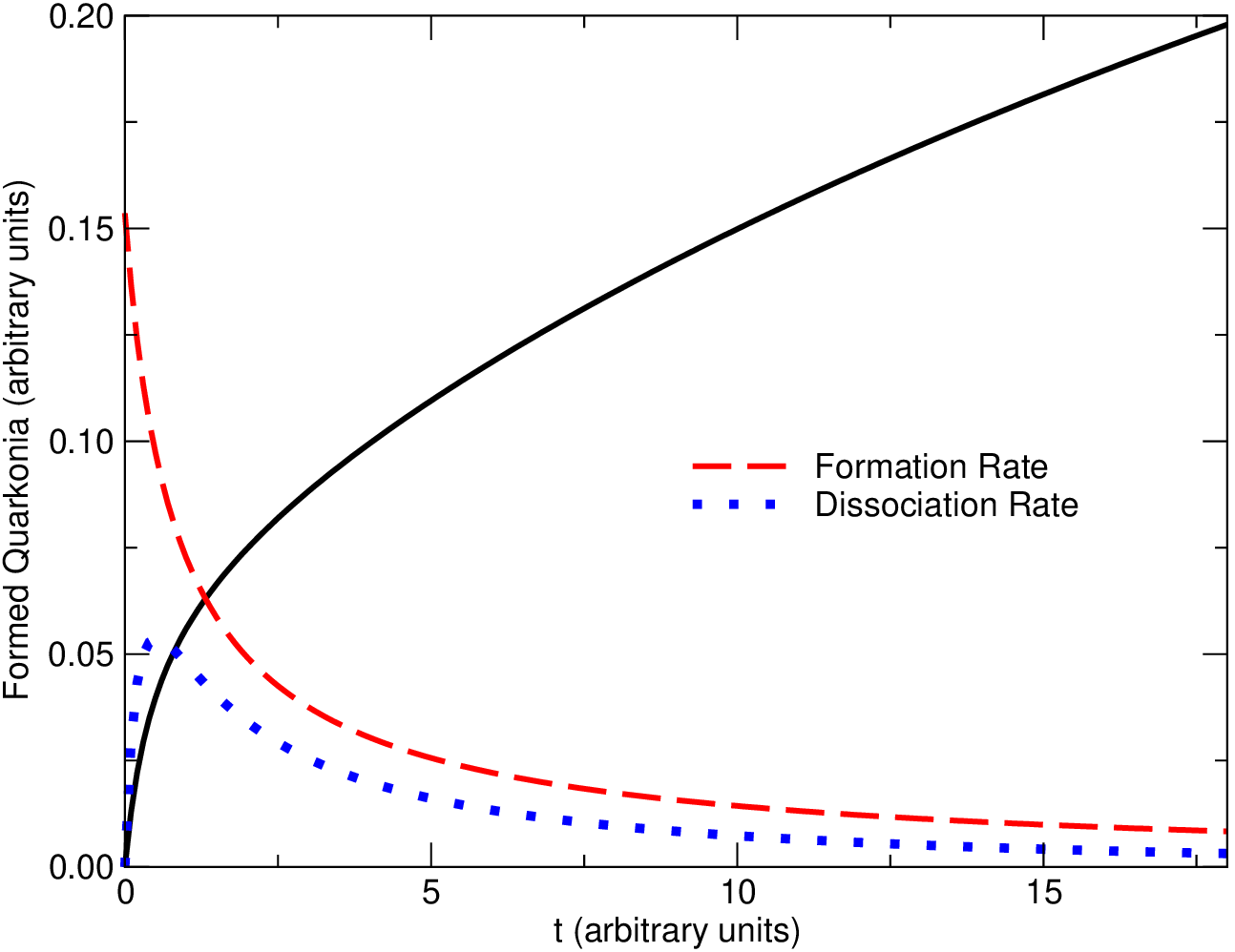}
}
\caption[]{Time development of $\J$ with formation and dissociation rates.
}
\label{formationtimedevelopment}
\end{figure}

Fig. \ref{formationtimedevelopment} shows the generic behavior of the 
$\J$ population resulting from a numerical solution of Equation \ref{eqkin}.
One can see that the final population is in fact determined by
the time integral of the difference between
formation and dissociation rates, shown as dashed lines.  The magnitudes are 
determined by the parameter $T_0$, which controls the magnitude and time 
dependence of the gluon density, and also the total lifetime.  It is important
to note that for parameter values in a range consistent with expectations, 
the expansion rate of the color deconfinement volume and the 
decrease of gluon density with time prevent the system from reaching an
equilibrium population within this lifetime.  

\begin{figure}
\vskip 1.0cm
\resizebox{0.45\textwidth}{!}{%
  \includegraphics{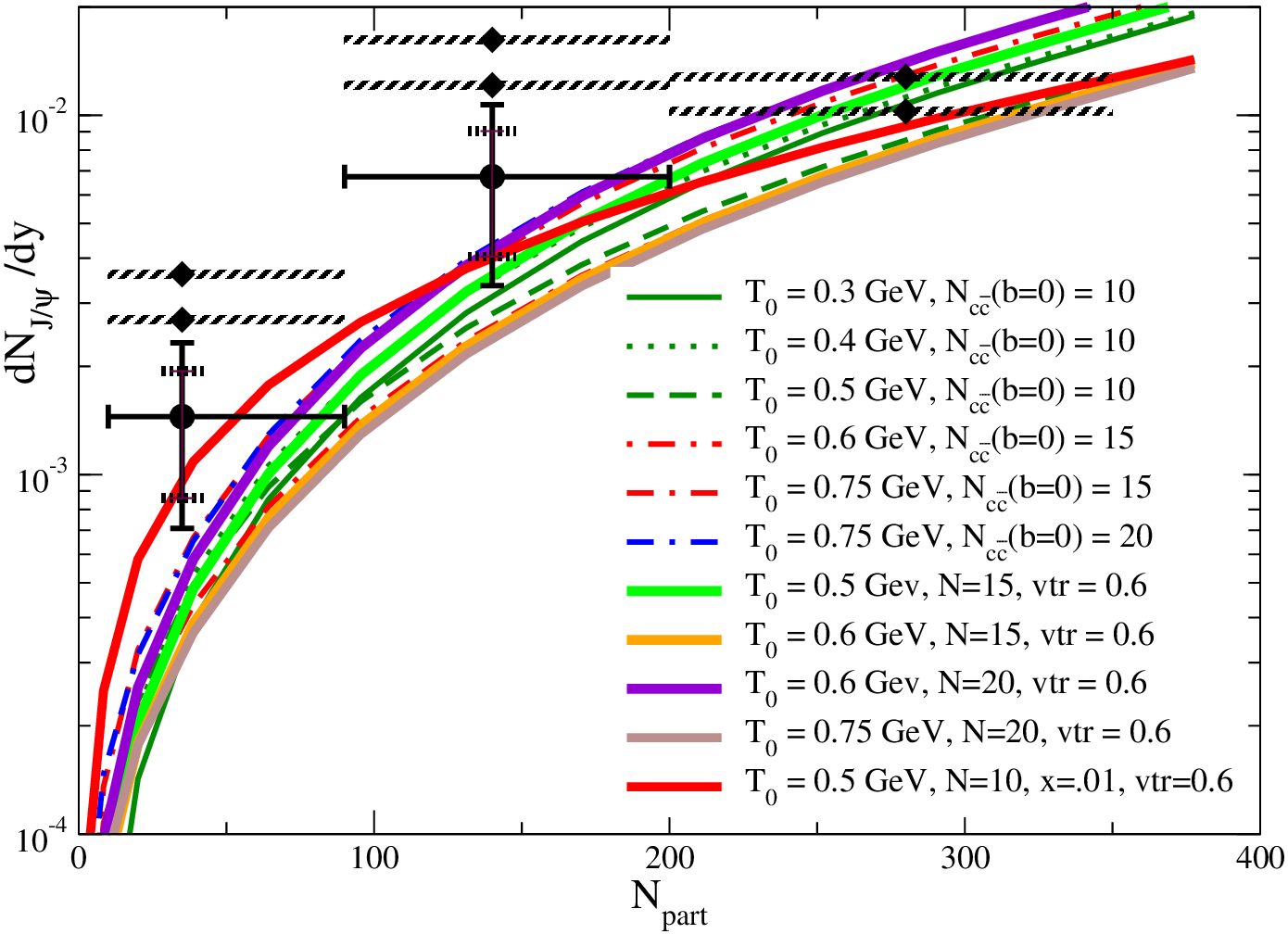}
}
\caption[]{$\J$ formation magnitudes and centrality dependence for 200 GeV
Au-Au at RHIC.
}
\label{rhicjpsidndy}
\end{figure}

Fig. \ref{rhicjpsidndy} shows the PHENIX data for $\J$ production in Au-Au
collisions at 200 Gev \cite{phenixjpsi}.  The most central bin yield only
allowed an upperlimit (hatched horizontal lines), while two less central
bins yielded absolute values plus additional one-sigma upper limits both 
from statistical and systematic uncertainties.  The lines shown are 
calculations in the kinetic model with a range of principal parameters, 
including $\Nccbar$ (b=0), $T_0$, transverse expansion velocity (vtr),
and in one case an initial population fraction (x) of $\J$.  The range of
these parameters was chosen to exhibit what constraints are placed by this
initial data.  All of the calculations used the same charm-quark distribution,
which was taken from a LO pQCD calculation \cite{mangano}.
One sees that there is a substantial range of parameters  allowed
by this data, but that the increase in statistics anticipated in run 3 
will allow a much more stringent limit for the acceptable (if any) region of
parameter space in the kinetic model.  

The kinetic model also makes predictions for the momentum space distribution of
formed $\J$.  For this purpose we require the differential cross sections
related to the  $\sigma_{F,D}$.  These are obtained via an adaptation of
the corresponding expressions  for photodissociation of atomic bound states.
One can then express the time-integrated formation rate in terms of a sum
over all $\ccbar$ pairs, each weighted by differential formation probabilities.

\begin{equation} 
{{dN_{\J}} \over d^3 P_{\J}} = \int{{dt} \over {V(t)}}
\sum_{i=1}^{N_c} \sum_{j=1}^{N_{\bar c}} {\it {v}_{rel}} 
{{d \sigma} \over d^3 P_{\J}}(P_c + P_{\bar{c}} \rightarrow P_{\J} + X)
\end{equation}
Note that the formation magnitude exhibits the explicit quadratic
dependence on total charm, normalized by a prefactor which 
is proportional to the inverse of the
system volume.

\begin{figure}
\vskip 1.0cm
\resizebox{0.45\textwidth}{!}{%
  \includegraphics{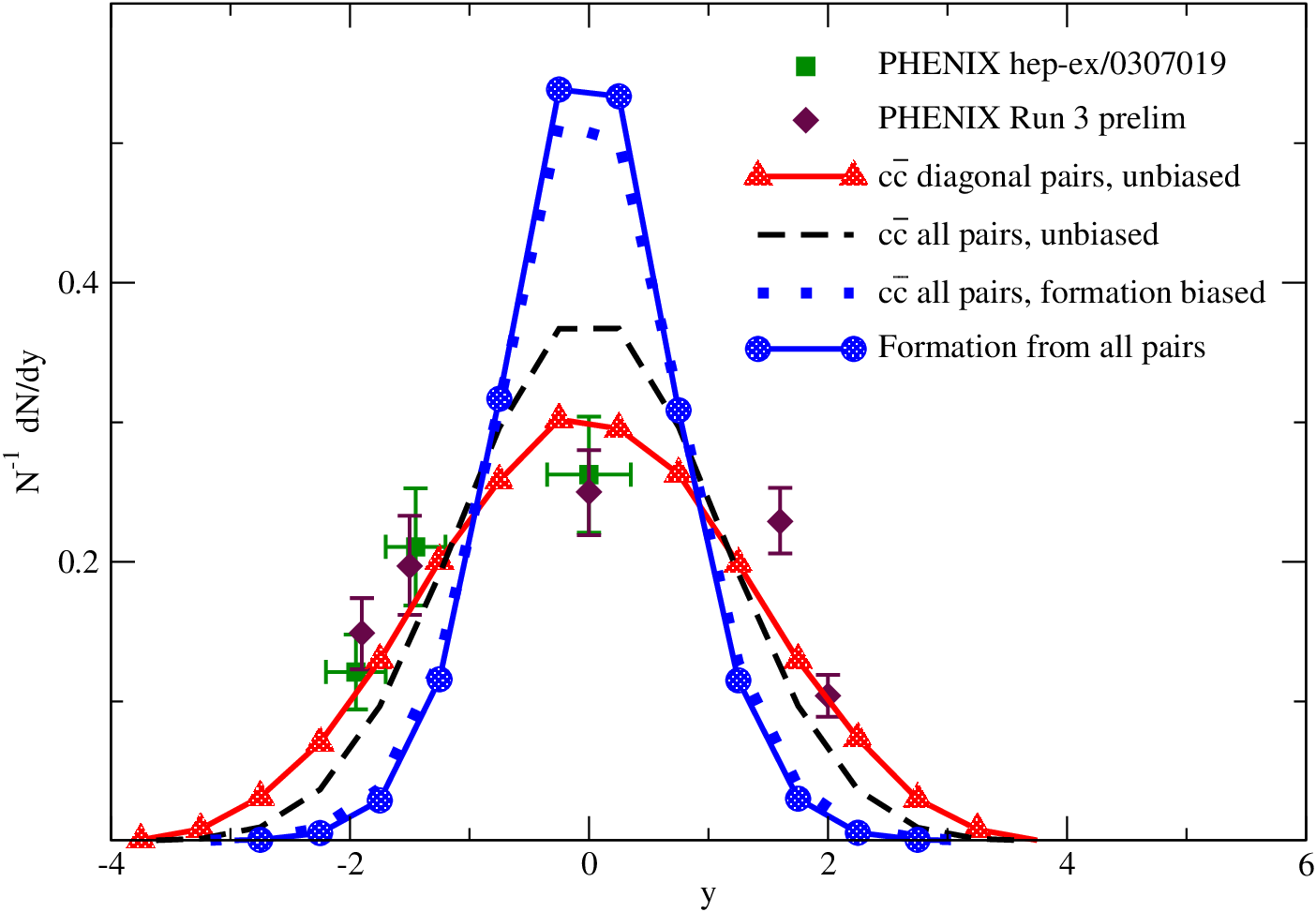}
}
\caption[]{Normalized rapidity spectra of c$\bar{c}$ pairs and $\J$
formation for 200 GeV Au-Au at RHIC.
}
\label{ccbarandjpsiyspectra}
\end{figure}

\begin{figure}
\vskip 1.0cm
\resizebox{0.45\textwidth}{!}{%
  \includegraphics{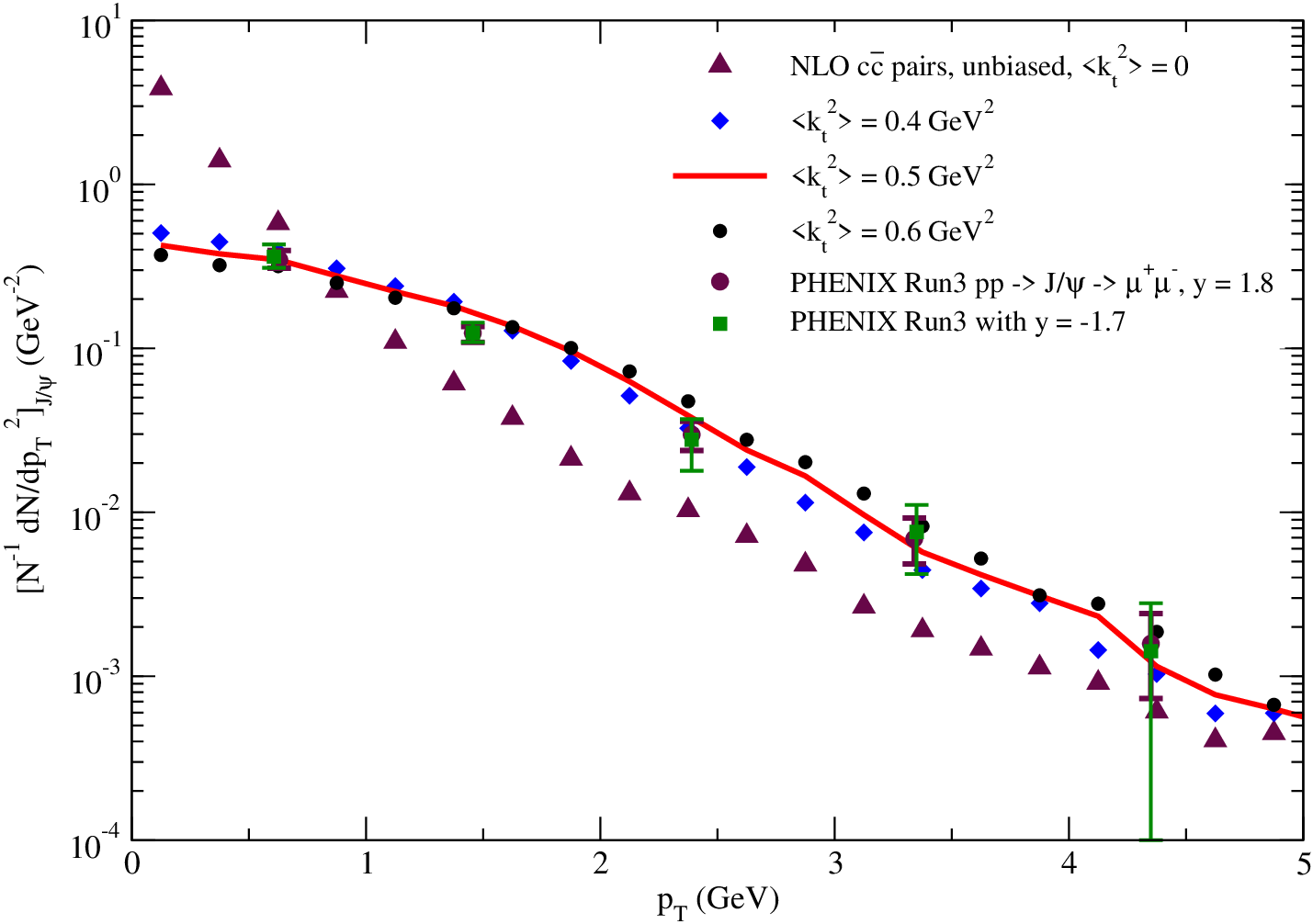}
}
\caption[]{Transverse momentum distribution for diagonal
c$\bar c$ pairs.}
\label{ppptspectra}
\end{figure}

\begin{figure}
\vskip 1.0cm
\resizebox{0.45\textwidth}{!}{%
  \includegraphics{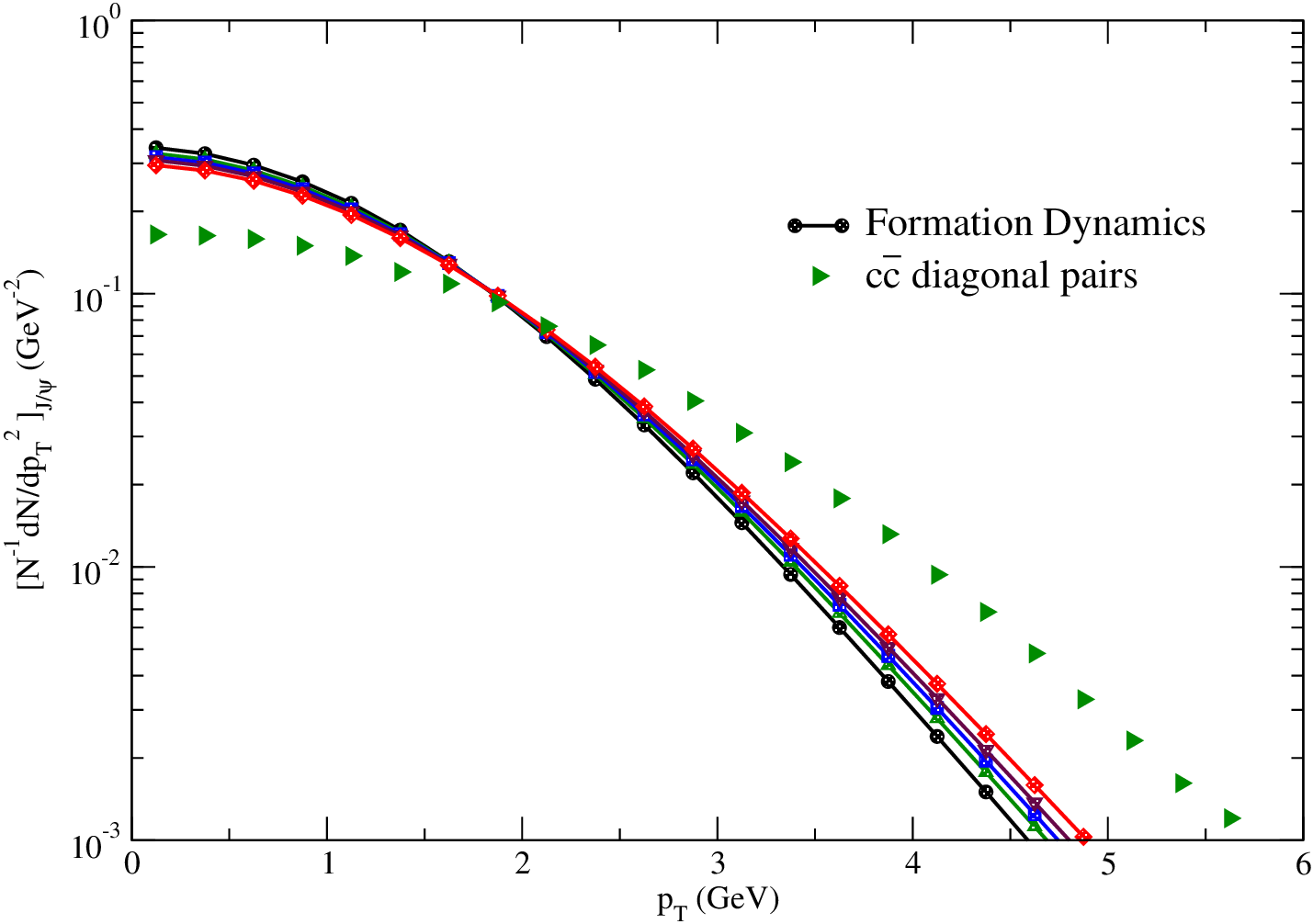}
}
\caption[]{Prediction for $\J$ $p_t$ distribution
from formation process in Au-Au collisions at RHIC.
}
\label{ptpredictions}
\end{figure}

We first look at the rapidity spectra of $\ccbar$ pairs, shown in
Fig. \ref{ccbarandjpsiyspectra}, and compare with the 
measured $\J$ distribution in pp interactions \cite{phenixpp}
. Normalized spectra
are used throughout, so that the results are independent of the
prefactors.   One sees that the data are consistent with the
distribution of unbiased diagonal pairs only, which is what one
would expect for pp interactions.  The distribution
of all pairs (also unbiased) is somewhat narrower than the data.  
Also shown are all pairs for which each is biased by the total
formation probability appropriate for the given pair energy.  It is
very close to the next curve, which takes into account 
formation from all pairs, using
exact $\J$ kinematics and the full differential dependence.  One sees
that both of these curves are substantially narrower than the pp
data. Thus the kinetic model predicts that 
the rapidity distribution of $\J$ formed by off-diagonal pairs (only
possible in nucleus-nucleus collisions which lead to color 
deconfinement) will be substantially narrower than $\J$ produced
in pp interactions at the same energy.
Fig. \ref{ppptspectra}
shows the transverse momentum spectra of 
unbiased diagonal $c \bar c$ pairs, along with the PHENIX 
data \cite{phenixpp} for $\J$ production in pp interactions at 200 GeV. 
The set of curves result from 
augmenting the quark initial momenta with a transverse momentum
``kick" to simulate confinement and initial state effects.  
The pp data restricts the magnitude of this kick, parameterized by
a Gaussian distribution, to lie 
within the range $<k_t^2>_{pp} = 0.5 \pm 0.1 GeV^2$.
To extend this to formation in Au-Au collisions, we must extract
the appropriate $k_t$ for initial state effects in the nucleus.
We use PHENIX data for $\J$ in d-Au collisions \cite{phenixdau}, 
which shows that
the $p_t$ spectra are broadened relative to that in pp interactions.
 This results in an estimate for
$<k_t^2>_{Au-Au} = 1.3 \pm 0.3 GeV^2$, where the uncertainty 
is set by the rapidity variation of the $\J$ $p_t$ broadening.  
This range of values was
utilized in the formation calculations in Au-Au interactions.
The predicted rapidity spectra are found to be essentially independent
of the magnitude of the initial charm quark $k_t$ kick, so that
the narrowest of the curves in Fig. \ref{ccbarandjpsiyspectra} 
will serve as
the kinetic model prediction for $\J$ formed in an Au-Au collision.
Fig. \ref{ptpredictions} shows the predicted transverse momentum
spectra of $\J$ at RHIC which would result from the formation
mechanism, using the entire allowed range of $<k_t^2>_{Au-Au}$.  
For comparison we show the distribution 
of diagonal unbiased $c \bar c$ pairs with the
central value in the allowed range of initial $<k_t^2>_{Au-Au}$, 
which should be relevant if all of the
$\J$ were produced directly from the initial $\ccbar$ pairs.
Of course, both of these distributions would be modified by
the competing dissociation process during the expansion phase,
but one would anticipate a similar effect on each which would
preserve the relative comparison.  (A sample suppression
factor applied to these curves actually shows very little
change in the shape of the {\it normalized} spectra.)

\section{Summary}

One expects on general grounds that heavy quarkonium production in
high energy heavy ion collisions must contain a component which
is formed either during a period of color deconfinement or at the
hadronization point.  The magnitude of this formation will
increase quadratically with the total amount of charm initially
produced via nucleon-nucleon interactions.  Both the
statistical hadronization model and the kinetic formation model
exhibit this property.  The absolute magnitude is somewhat
model-dependent, and initial RHIC data for $\J$ can be accommodated.
The rapidity and transverse momentum spectra may be decisive in
determining whether or not this formation makes a significant
contribution.  The kinetic formation model predicts a narrowing 
of the $\J$ rapidity distribution (compared with that in pp collisions),
and also a narrowing of the transverse momentum distribution 
(compared with an extrapolation of behavior measured in pp and
d-Au collisions).  In principle, both of these formation mechanisms
can coexist, so that the upcoming Au-Au data may reveal a
two-component structure.

\begin{acknowledgement}
This work was supported by U. S. Department of
Energy Grant DE-FG02-04ER41318.
\end{acknowledgement}

\end{document}